\documentclass[
superscriptaddress,
amsmath, amssymb,
aps,
prl,
dblfloatfix,
nobalancelastpage,
twocolumn
]{revtex4-1}
\usepackage{graphicx}
\usepackage{bm}
\usepackage{siunitx}
\usepackage{hyperref}
\usepackage{braket}
\usepackage{epstopdf}
\hypersetup{colorlinks=true, citecolor=blue, urlcolor=blue, linkcolor=blue}

\begin{document}

\preprint{APS/123-QED}

\title{Exact Higher-order Bulk-boundary Correspondence of Corner-localized States}

\author{Minwoo Jung}
\email{mj397@cornell.edu}
\affiliation{Department of Physics, Cornell University, Ithaca, New York, 14853, USA}
\author{Yang Yu}
\author{Gennady Shvets}
\email{gshvets@cornell.edu}
\affiliation{School of Applied and Engineering Physics, Cornell University, Ithaca, New York 14853, USA}

\date{\today}

\begin{abstract}
We demonstrate that the presence of a localized state at the corner of an insulating domain is not always a predictor of a certain non-trivial higher-order topological invariant, even though they appear to co-exist in the same Hamiltonian parameter space. Our analysis of $C^n$-symmetric crystalline insulators and their multi-layer stacks reveals that topological corner states are not necessarily correlated with other well-established higher-order boundary observables, such as fractional corner charge or filling anomaly. In a $C^3$-symmetric breathing Kagome lattice, for example, we show that the bulk polarization, which successfully predicts the fractional corner anomaly, fails to be the relevant topological invariant for zero-energy corner states; instead, these corner states can be exactly explained by the decoration of topological edges. Also, while the zero-energy corner states in $C^4$-symmetric topological crystalline insulators have long been conjectured to be the result of the bulk polarization at quarter-filling, we correct this misconception by introducing a proper bulk invariant at half-filling and establishing a precise bulk-corner correspondence. By refining several bulk-corner correspondences in two-dimensional topological crystalline insulators, our work motivates further development of rigorous theoretical grounds for associating the existence of corner states with higher-order topology of host materials.
\end{abstract}

\maketitle
\section{I. Introduction}
Bulk-boundary correspondence (BBC) lies at the heart of topological physics. BBC bridges abstract mathematical indices called topological invariants, which are calculated from band structures of a bulk material, to physical observables at its boundary. Early efforts in establishing BBC focused on boundaries of co-dimension 1 such as edges of two-dimensional (2D) materials or surfaces of three-dimensional (3D) materials \cite{BBC_1st,BBC_2nd,BBC_3rd,BBC_4th,BBC_5th}. Inspired by the discovery---both theoretical\cite{HOTI_theory_1st, HOTI_theory_2nd, HOTI_theory_3rd, HOTI_theory_4th, HOTI_theory_5th} and experimental\cite{HOTI_exp_1st, HOTI_exp_2nd, HOTI_exp_3rd, HOTI_exp_4th, HOTI_exp_5th}---of topological materials that feature gapless states at boundaries of co-dimension $d\geq 2$, efforts have recently been made to extend the framework of BBC to these higher-order topological phases \cite{HOBBC_1st, HOBBC_2nd, HOBBC_3rd}. 

The study of BBC sometimes takes a form of analytic case studies with a specific form of topological invariant \cite{BBC_3rd, BBC_4th, HOBBC_3rd}, or relies on algebraic topology for generic classification of bulk and boundary Hamiltonians \cite{BBC_1st,BBC_2nd,BBC_5th, HOBBC_1st, HOBBC_2nd}. While the latter approach provides more comprehensive formulation of BBC than the former does, its concern does not aim further than identifying the classification group of Hamiltonian in certain symmetry classes, thereby evading the task of finding the actual topological invariants relevant to the boundary signatures. Therefore, while the algebraic classification method allows an insightful start for the search of topological structures, rigorous BBC cannot be established without rigorous case studies. 

As the field of higher-order topological insulators (HOTIs) rapidly expanded, however, rigorous BBCs have often been replaced by an implicit assumption that the boundary signatures (e.g., corner-localized states) must be related to a specific bulk invariant that the host bulk Hamiltonian is most famously characterized by. To be specific, the following prescriptive framework is widely used in the field of HOTIs: (1) find a symmetry-protected bulk topological invariant of a given Hamiltonian model within a certain range of parameters, (2) uncover corner-localized states for the same parameters of the Hamiltonian as in (1), and (3) conflate (1) and (2) because both occur for the same parameters range. Because correlation does not imply causation, the above procedure does not necessarily imply that the boundary signature has a topological origin.  A physical explanation is necessary to establish the causal relationship between bulk invariants and the emergence of anomalous boundary properties. Otherwise, a topological nature of a boundary can be attributed to an irrelevant bulk invariant, or a trivial defect state can be mistaken for a topological one, thereby obscuring true BBCs.

In this work, we address several cases of such weakly conjectured higher-order BBC, specifically in the context of corner-localized states and bulk polarization in 2D $C^n$-symmetric topological crystalline insulators (TCIs). We reveal that the bulk polarization is not the right invariant to be associated with the emergence of the zero-energy corner states (ZCSs) in 2D TCIs. The multilayer stacking construction of TCIs \cite{Bena_PRB_2019} reveals that ZCSs in $C^3$-symmetric TCI (also known as breathing Kagome lattices) \cite{FCA, Bena_PRB_2019, Kagome_1st, Kagome_2nd, Kagome_3rd} are purely an edge effect associated with $\mathbb{Z}_2$ composite Zak-phase of chiral-symmetric edge bands and are not well correlated with the bulk polarization characterized by $\mathbb{Z}_3$. Also, we discuss that the ZCSs in 4-band $C^4$-symmetric TCI model \cite{C4_BIC, C4_BIC2, FCA, Bena_PRB_2019} should not be attributed to the bulk polarization of the lowest energy band, while all previous studies regarding this $C^4$-symmetric TCI model \cite{C4_BIC, C4_BIC2, FCA, C4_wrong_1st, C4_wrong_2nd} conjectured such imprecise BBC between the ZCSs and the bulk polarization of the lowest energy band. Instead, it turns out that the corner charge index defined at half-filling (the first and the second band altogether) instead is a proper bulk invariant responsible for ZCSs. These examples clearly demonstrate that a precise formulation of BBC requires more than simply identifying the phase diagrams of a bulk invariant and a boundary state.

\begin{figure}[t]
\centering
    \includegraphics[width=0.95\columnwidth]{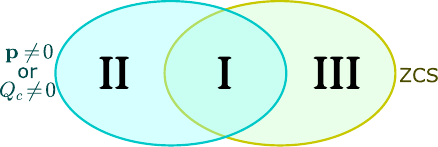}
    \caption{A Venn diagram classification of crystalline insulators in regard to the existence of nonzero $\bf{p}$ or $Q_c$ and the existence of ZCSs; $\bf{I}$ : a class of models that support both nontrivial bulk higher-order topology and ZCSs, $\bf{II}$ : support nonzero $\bf{p}$ or $Q_c$ only, but no ZCSs, and $\bf{III}$ : support ZCSs despite trivial bulk topology.}
    \label{fig.1.}
\end{figure}

\begin{figure}[b]
\centering
    \includegraphics[width=0.95\columnwidth]{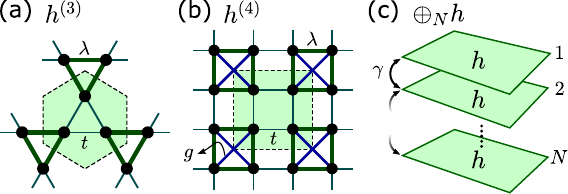}
    \caption{(a) $C^3$-symmetric crystalline insulator model $h^{(3)}$; $t$ and $\lambda$ are the nearest-neighbor coupling within and across unit cells, respectively. (b) $C^4$-symmetric crystalline insulator model $h^{(4)}$; $g$ is the next-nearest-neighbor coupling across diagonally adjacent unit cells. (c) $N$-layer stacking of a Hamiltonian model $h$, $\oplus_N h$; $\gamma$ denotes the interlayer coupling strength.}
    \label{fig.2.}
\end{figure}

To set some good examples, we briefly review several well-established BBCs. A classic example is the correspondence between the Hall conductivity (a surface effect) and Chern number (a bulk topological index). The two are directly related through an analytic expression~\cite{TKNN}. Because conduction cannot occur in an insulating bulk, nonzero Hall conductivity in a Chern insulator must indicate metallic channels on its edge or, in other words, gapless edge states \cite{TKNN2}. Another analytically straightforward BBC is found between fractional edge charge and bulk polarization in a one-dimensional (1D) TCI \cite{Edge_charge_1D}, i.e. Su-Schrieffer-Heeger (SSH) model. Recent work \cite{Bena_PRB_2019, Bena_PRB_2017} established higher-order versions of similar correspondences in 2D TCIs by explicitly constructing bulk invariants  for fractional corner charges. Even though these boundary anomalies in the form of fractional charge excess/deficit have yet to be incorporated in the framework of the algebraic classification method \cite{HOBBC_2nd}, they have recently attracted attention as alternative probes of higher-order topology \cite{FCA}.

Our key result is that nontrivial bulk polarization $\bf{p}$ or secondary topological index for corner charge $Q_{\rm{c}}$ \cite{Bena_PRB_2019} (and fractional corner or edge charge anomalies) are not always sufficient or necessary for the existence of ZCSs, even though they appear to co-exist in some systems. This result is starkly different from the above-mentioned non-vanishing Chern number (and non-zero Hall conductivity) being a necessary and sufficient condition for the existence of gapless edge states. The Venn diagram shown in Fig.\ref{fig.1.} schematically illustrates our key result: a ZCS might not exist despite nontrivial $\bf{p}$ and $Q_{\rm{c}}$ (classification set II) or a ZCS can arise despite vanishing $\bf{p}$ and $Q_{\rm{c}}$ (classification set III). Figure \ref{fig.2.}(a) and (b) depict $C^3$- and $C^4$-symmetric crystalline insulators, respectively, that are used as exemplary models to support our key results, and Table \ref{tab1} summarizes various $C^3$- and $C^4$-symmetric Hamiltonian models according to each classification category defined in Fig.\ref{fig.1.}. Each Hamiltonian model elements in Table \ref{tab1} is discussed in details in the following sections.

\begin{table}
\caption{\label{tab1}Hamiltonian models of $C^3$- and $C^4$-symmetric insulators for each classification set introduced in Fig. \ref{fig.1.}.}
\begin{ruledtabular}
\begin{tabular}{p{0.5in} p{1.5in} p{1.5in}}
 & $C^3$-symm. models & $C^4$-symm. models \\
\hline
I & $\oplus_{1,5,7,11,...} h^{(3)}(|t|<|\lambda|)$ & $h^{(4)}(|t|<|\lambda|;g=0)$ \\
II & $\oplus_{2,4,8,10,...} h^{(3)}(|t|<|\lambda|)$ & $h^{(4)}(|t|<|\lambda|;g\neq0)$ \\
III & $\oplus_{3,9,15,...} h^{(3)}(|t|<|\lambda|)$ & $h^{(4)}(|t|>|g|;\lambda=0)$ \\
\end{tabular}
\end{ruledtabular}
\end{table}

\section{II. Stacking Operation}
We introduce the stacking operation $\oplus$ between two crystalline insulators $h_1$ and $h_2$, as defined in Ref.\cite{Bena_PRB_2019}
\begin{equation}
h_1 \oplus h_2 = \begin{bmatrix} h_1 & \gamma \\ \gamma^{\dagger} & h_2\end{bmatrix}, \label{eq1}
\end{equation}
where $\gamma$ describes the nearest-neighbor coupling between adjacent layers. The strength of interlayer coupling is set to be reasonably small so that the shared bandgap of $h_{1,2}$ is not closed. We denote an $N$ layer stack of $h$ as $\oplus_{N} h$, as depicted in Fig. \ref{fig.2.}(c). This operation allows us to easily access other topologically distinct phases, as the topological indices of a stacked insulator are simply given as addition of those in each layer \cite{Bena_PRB_2019}; for example,
\begin{equation}
\bf{p}_{\it{h}_{\rm{1}} \oplus \it{h}_{\rm{2}}} \equiv \bf{p}_{\it{h}_{\rm{1}}} + \bf{p}_{\it{h}_{\rm{2}}} \; (\rm{mod}\, \left\lbrace \bf{R}\right\rbrace), \label{eq2}
\end{equation}
where the composite polarization $\bf{p}$ (normalized to a unit charge) is evaluated in each model for all bands below the shared bandgap of interest, and is given in modulo the set of primitive lattice vectors $\left\lbrace\bf{R}\right\rbrace$. The same relation holds for $Q_c$ as well in modulo unit charge. The stacking operation defined here in Eq. (\ref{eq1}) and its property in Eq. (\ref{eq2}) turn out to be extremely useful in constructing case models that belong to each category of the Venn diagram in Fig. \ref{fig.1.}, especially for $C^3$-symmetric crystalline insulators as shown in Table \ref{tab1}. 

\section{III. ZCS in $C^3$ TCI as boundary topological effects}
\begin{figure}[t]
\centering
    \includegraphics[width=0.95\columnwidth]{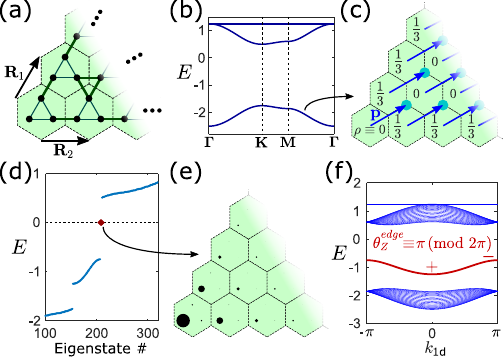}
    \caption{(a) A $60^{\circ}$-angled corner with a single corner-most sublattice. (b) Band structure of $h^{(3)}$; $t=-0.25$ and $\lambda=-1$. (c) Charge distribution around a $60^{\circ}$-angled corner at $\frac{1}{3}$-filling (upto the first band only); turquoise circles denote the Wannier centers displaced by $\bf{p} = \rm\frac{2}{3}\bf{R}_{\rm1}+\rm\frac{2}{3}\bf{R}_{\rm2}$ from the unit cell centers. (d) Eigenspectra of a finite-sized system (190 unit cells) with open boundaries of triangular termination like in (a); corner-localized modes are highlighted as dark red dots. (e) Field profile of a ZCS; the area of black circles are proportional to the wavefunction amplitude. (f) Edge dispersion of nano-ribbon structure with an edge termination like one of the edges in (a); the edge localized band is colored red along with its inversion eigenvalues at high symmetry points.}
    \label{fig.3.}
\end{figure}

When the nearest-neighbor coupling strengths across unit cells, $\lambda$, are greater than those within unit cells, $t$, $C^3$-symmetric model $h^{(3)}$ in Fig. \ref{fig.2.}(a), also known as a breating Kagome lattice, carries a ZCS emerging at every $60^{\circ}$-angled corners of a type with a single corner-most sublattice, as depicted in Fig. \ref{fig.3.}(a) (another type of $60^{\circ}$-angled corners with two corner-most sublattices doesn't support ZCSs). The same condition $|t|<|\lambda|$ produces nonzero bulk polarization $\bf{p}_{\rm{(1)}} = \rm\frac{2}{3}\bf{R}_{\rm1}+\rm\frac{2}{3}\bf{R}_{\rm2}$ in the lowest energy band \cite{Bena_PRB_2019, Kagome_1st, Kagome_2nd, Kagome_3rd}, which is separated from the second and third bands by a bandgap, see Fig. \ref{fig.3.}(b). Figure \ref{fig.3.}(c) illustrates that each Wannier center is displaced from the origin of each unit cell by bulk polarization vector $\bf{p}_{\rm{(1)}}$, and therefore located at the junction vertice of three adjacent hexagonal unit cells. Thus, in the limit of $|t|\ll|\lambda|$ (i.e. localization length of wannier function is much smaller than the unit cell size), a unit cell gains fractional charge of $\frac{1}{3}$ from each Wannier center in contact, when the lowest energy band is occupied. For example, the corner-most unit cell carries no charge $\rho=0$ as there is no Wannier center in contact, each unit cell along both edges carries a fractional charge of $\rho=\frac{1}{3}$ as there is a Wannier center in contact, and each unit cell in the bulk carry a whole charge $\rho\equiv 0$ (mod 1) as there are three Wannier centers in contact. From this observation, it has been recently established that nonzero bulk polarization in $h^{(3)}$ gives rise to a higher-order topological observable called fractional corner anomaly (FCA) $\phi=\rho_{\text{corner}} -\rho_{\text{edge1}} - \rho_{\text{edge2}} = 0-\frac{1}{3}-\frac{1}{3}\equiv\frac{1}{3}$ (mod 1)\cite{FCA}. Note that FCA is non-vanishing even in the absence of corner charge $\rho_{corner}=Q_c$.

Therefore, it may be tempting to conclude that the existence of a ZCS shown in Fig.\ref{fig.3.} (d) and (e) is correlated with either a finite $Q_c$, or at least with a finite FCA. This conjecture is disproved by our analysis of multi-layer stacking constructions of $\oplus_{1,2,3...}h^{(3)}\left(|t|<|\lambda|\right)$ described below. Instead, we prove that the existence of a ZCS is a result of topological Zak phase of the edge localized band. Figure \ref{fig.3.}(f) shows the band dispersion of 1D-periodic nano-ribbon structure terminated by an edge shown in Fig \ref{fig.3.}(a), where the red line denotes an edge-localized band. This edge band carries the inversion eigenvalues of $+1$ at $k_{1d}=0$ and $-1$ at $k_{1d}=\pi$, thereby featuring a Zak phase of $\theta_Z^{\text{edge}}=\pi$ \cite{Edge_charge_1D} (or polarization of $\frac{1}{2}$ \cite{Bena_PRB_2017}). The energy dispersion of this band follows $E(k_{1d})=-\sqrt{t^2 + \lambda^2 + 2t\lambda\cos(k_{1d})}$, which is reminiscent of a 1D chiral-symmetric SSH chain \cite{Edge_charge_1D}. In fact, the chiral partner band of this edge band in Fig. \ref{fig.3.}(d) does not stand out since it is hybridized with other bulk bands at positive energy. The detailed discussion on how this edge-localized band is exactly mapped onto a 1D chiral-symmetric SSH chain is provided in Appendix.

\begin{figure}[t]
\centering
    \includegraphics[width=0.95\columnwidth]{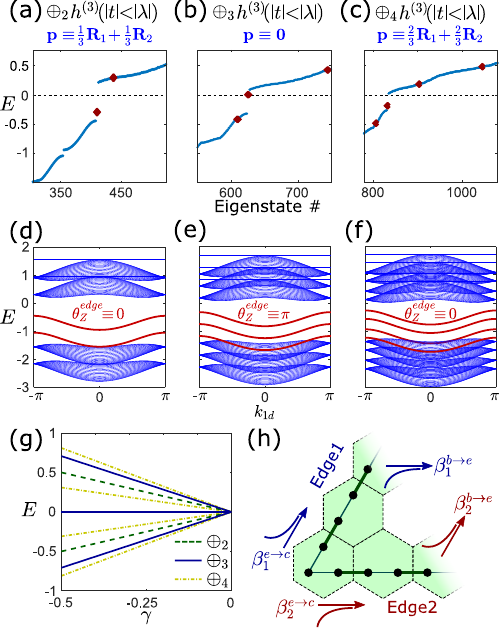}
    \caption{(a)-(c) Eigenspectra of a finite-sized system---190 unit cells as in Fig. \ref{fig.3.}(d)---of the bi-, tri-, and quad-layer stacked structures of $h^{(3)}$($t=-0.25$, $\lambda=-1$), respectively, with the interlayer coupling strength of $\gamma=-0.3$; corner-localized modes are highlighted as dark red dots. (d)-(f) Edge dispersion of nano-ribbon structure of bi-, tri-, and quad-layer structures, respectively, as in Fig. \ref{fig.3.}(f); the composite Zak phases of the edge-localized bands (dark red) are denoted together. (g) Energies of all corner-localized states in bi-(dashed green), tri-(solid blue), and quad-layer(dot/dashed orange) structures as a function of $\gamma$. (h) At the corner, each of two edge-localized SSH chains ($i=1,2$) supports a zero-energy termination-localized state, where it is localized along the edge with the edge-to-corner localization factor $\beta^{e\rightarrow c}_i$ and localized with respect to the bulk with the bulk-to-edge localization factor $\beta^{b\rightarrow e}_i$. These two states coalesce to each other as $\beta^{b\rightarrow e}_1=\beta^{e\rightarrow c}_2$ and $\beta^{b\rightarrow e}_2=\beta^{e\rightarrow c}_1$, giving rise to a ZCS to the bulk Hamiltonian.}
    \label{fig.4.}
\end{figure}

In what follows, we analyze the multi-layer stacks of $h^{(3)}\left(|t|<|\lambda|\right)$ to show that the existence of ZCSs of a breathing Kagome lattice is correlated with neither finite bulk polarization nor with finite FCA. Such correlation has been widely assumed because the existence conditions $|t|<|\lambda|$ for ZCSs and nonzero $\bf{p}$ appear to coincide with each other \cite{Kagome_1st, Kagome_2nd, Kagome_3rd}. Bilayer and trilayer stacks of $h^{(3)}\left(|t|<|\lambda|\right)$, according to Eq. (\ref{eq2}), carry the bulk polarization of $\bf{p}_{\rm{(2)}} \equiv \rm 2 \bf{p}_{\rm{(1)}}\equiv \rm\frac{1}{3}\bf{R}_{\rm1}+\rm\frac{1}{3}\bf{R}_{\rm2}$ and $\bf{p}_{\rm{(3)}} \equiv \rm 3 \bf{p}_{\rm{(1)}}\equiv 0$, respectively. Thus, based on their bulk polarization, $\oplus_2 h^{(3)}$ is classified as topologically nontrivial, and $\oplus_3 h^{(3)}$ as trivial. This distinction will indeed physically manifest in their FCA; $\phi=\frac{2}{3}$ for $\oplus_2 h^{(3)}$ and $\phi=0$ for $\oplus_3 h^{(3)}$. Therefore, if the presence of a ZCS were predicated on the finite FCA, we would expect that $\oplus_2 h^{(3)}$ should posses a ZCS while $\oplus_3 h^{(3)}$ should not. Remarkably, the opposite is true, as observed from Fig.~\ref{fig.4.}(a-b). Furthermore, the quad-layer stack $\oplus_4 h^{(3)}$ shares exactly the same bulk polarization $\bf{p}_{\rm{(4)}} \equiv \rm 4 \bf{p}_{\rm{(1)}}\equiv \bf{p}_{\rm{(1)}}$ and FCA $\phi=\frac{1}{3}$ with the original monolayer structure $h^{(3)}$ that supports ZCSs, but $\oplus_4 h^{(3)}$ does not support a ZCS as shown in Fig.~\ref{fig.4.}(c).

On the other hand, the composite Zak phase of the edge-localized bands in those structures, as shown in Fig. \ref{fig.4.}(d)-(f), predicts well the existence of ZCSs. In the presence of multiple bands below a certain bandgap of interest, the existence of a mid-gap boundary/dislocation state in 1D systems is determined by the composite Zak phase of all bands below the bandgap \cite{CompositeZak1, CompositeZak2}. Thus, we find that a stack with an even number of layers features vanishing $\theta_Z^{edge}=0$ (mod $2\pi$) and a stack with an odd number of layer has nontrivial $\theta_Z^{edge}=\pi$. Accordingly, we observe the ZCSs in odd-layer stacks, but not in even-layer stacks. We note that there exist two corner-localized states in the bilayer stack structure as well, but they are not pinned at zero-energy. Their spectral positions are at $E=\pm\gamma$, where $\gamma$ is the interlayer coupling strength. Consequently, these corner states are not spectrally stable against perturbations in $\gamma$ (e.g. vertical compression). Similarly, the trilayer stack also carries two spectrally unstable corner states at $E=\pm\sqrt{2}\gamma$ other than the ZCS. The spectral shifts of these corner states with respect to the change in $\gamma$ is drawn in Fig. \ref{fig.4.}(g). 

In general, $\oplus_{N}h^{(3)}\left(|t|<|\lambda|\right)$ carries $N$ corner-localized states, and, one of them becomes a ZCS with topological spectral pinning, when $N$ is an odd number. Therefore, it is clear that the existence of ZCSs is determined not by $\mathbb{Z}_3$ bulk polarization, but by $\mathbb{Z}_2$ edge band Zak phase. To be specific, a corner acts as a termination to each of two edge-localized SSH chains, and each topological ($\theta_Z^{edge} = \pi$) SSH chain is expected to support a zero-energy state localized at the termination: $\ket{v_{1}}=\sum_{n,m\geq0} \left[\beta^{b\rightarrow e}_1\right]^n \left[\beta^{e\rightarrow c}_1\right]^m \ket{A;n\bf{R}_{\rm 1}+\it m\bf{R}_{\rm 2}}$ and $\ket{v_{2}}=\sum_{n,m\geq0} \left[\beta^{b\rightarrow e}_2\right]^n \left[\beta^{e\rightarrow c}_2\right]^m \ket{A;n\bf{R}_{\rm 2}+\it m\bf{R}_{\rm 1}}$. Here, $A$ is the sublattice index of the corner-most sublattice, $\ket{A;\bf{R}}$ is the basis vector that occupies the sublattice $A$ in the unit cell located at position $\bf{R}$, and $\beta^{b\rightarrow e}_i$/$\beta^{e\rightarrow c}_i$ is the bulk-to-edge/edge-to-corner localization factor as depicted in Fig. \ref{fig.4.}(h). It turns out that these two localized states from each edge coalesce $\ket{v_{1}}=\ket{v_{2}}$, as the bulk-to-edge localization factor of an edge matches exactly to the edge-to-corner localization factor of the other edge: $\beta^{b\rightarrow e}_1 = \beta^{e\rightarrow c}_2=-t/\lambda$ and $\beta^{b\rightarrow e}_2 = \beta^{e\rightarrow c}_1=-t/\lambda$.

\begin{figure}[t]
\centering
    \includegraphics[width=0.95\columnwidth]{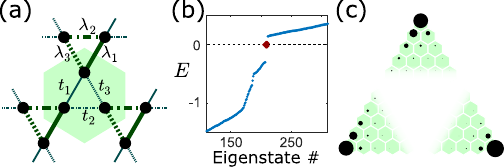}
    \caption{(a) Breathing Kagome lattice without any bulk crystalline ($C^3$- nor mirror) symmetries. (b) Eigenspectra of a finite-sized system---190 unit cells as in Fig. \ref{fig.3.}(d)---with $t_1=-0.5$, $\lambda_1=-0.9$, $t_2=-0.3$, $\lambda_2=-1$, $t_3=-0.1$, and $\lambda_3=-0.4$. (c) Field profiles of ZCSs at each corner.}
    \label{fig.5.}
\end{figure}

Lastly, we show that ZCSs still arise in a breathing Kagome lattice, when there is no bulk crystalline symmetry. Figure \ref{fig.5.} clearly demonstrates that ZCSs are well preserved even though hopping strengths are all different for three sides and $C^3$-rotational and mirror symmetries are broken. Like this case where we observe edge-induced corner states without any connection to bulk properties, several recent works have similarly identified higher-order topological signatures stemming from boundary (not bulk) topology in the language of decoration subgroups \cite{HOBBC_2nd} or embedded topological insulators \cite{Embedded}.

\begin{figure}[t]
\centering
    \includegraphics[width=0.95\columnwidth]{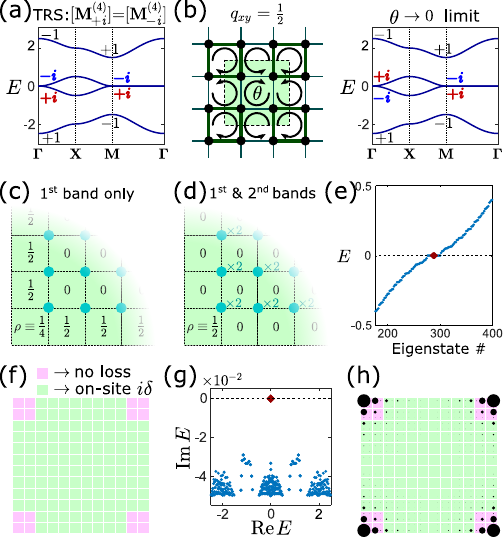}
    \caption{(a) Band structure of $h^{(4)}$ along with $C^4$-rotation eigenvalues at $\bold{\Gamma}$ and $\bold{M}$ ; $t=-0.25$, $\lambda=-1$, and $g=0$. (b) Staggering phase flux of $\theta$ realizes a quadrupole insulator, and the same structure in (a) viewed as a zero-flux limit ($\theta\rightarrow0$) of quadrupole insulator; the $C^4$-rotation eigenvalues at $\bold{\Gamma}$ and $\bold{M}$ for the second and third bands are assigned differently. (c) Charge distribution around a corner at quarter-filling (the first band only); turquoise circles are Wannier centers. (d) Charge distribution around a corner at half-filling (upto the second band); two wannier centers are overlapping at each position. (e) Eigenspectra of a finite-sized system with open boundaries (12 by 12 unit cells); the corner modes are marked as dark red dots. (f) A uniform on-site loss of $\delta=-0.05$ was applied to a open-boundary domain of $12\times12$ unit cells except at each $2\times2$ corner unit cells. (g) Resulting eigenvalues in a complex energy plane. (h) Field profile of the corner modes marked as dark red dots in (f) and (h).}
    \label{fig.6.}
\end{figure}

\section{IV. Well-defined corner charge index in gapless $C^4$ TCI at half-filling}
Next, we establish the correct BBC for the ZCS in $C^4$-symmetric TCI model, $h^{(4)}\left(|t|<|\lambda|;g=0\right)$ from Fig. \ref{fig.1.}(d). As briefly mentioned in the introduction, all of the previous works that studied the corner states in this $C^4$-symmetric TCI model have conjectured that their observation of the corner state must be a manifestation of the nonzero bulk polarization of the lowest band \cite{C4_BIC, C4_BIC2, FCA, C4_wrong_1st, C4_wrong_2nd}. Here, however, we show that the ZCS in this system is a result of a half corner charge from the lower two bands in the presence of chiral symmetry, not a result of the bulk polarization from the lowest band only.

First, we demonstrate that a half corner charge $Q_c=\frac{1}{2}$ is well defined at half-filling for the first and the second band considered together, even though the system is gapless at zero energy between the second and the third band. In a time reversal- and $C^4$-symmetric crystalline insulator, the following expressions can be used to determine its topological indices \cite{Bena_PRB_2019}:
\begin{subequations}
\label{eq3}
\begin{gather}
\bold{p} = \frac{1}{2}\left[\bold{X}_{+1}^{(2)}\right]\left(\hat{\bold{x}}+\hat{\bold{y}}\right) \; (\rm{mod}\, \left\lbrace \hat{\bold{x}},\hat{\bold{y}}\right\rbrace), \label{eq3a}\\
Q_c = \frac{1}{4}\left(\left[\bold{X}_{+1}^{(2)}\right]+2\left[\bold{M}_{+1}^{(4)}\right]+3\left[\bold{M}_{+i}^{(4)}\right]\right) \; (\rm{mod}\,1), \label{eq3b}
\end{gather}
\end{subequations}
where $[\bold{k}_{p}^{(n)}]\equiv \#\bold{k}_{p}^{(n)}-\#\bold{\Gamma}_{p}^{(n)}$, and $\#\bold{k}_{p}^{(n)}$ refers to the number of eigenstates with $C^n$-rotation eigenvalue $p$ at a $C^n$-rotational invariant momentum $\bold{k}$. The eigenstates are counted from the lowest propagation band up to the band of interest. For example, in $h^{(4)}\left(|t|<|\lambda|;g=0\right)$, the $C^2$-rotation eigenvalues are $(-1, +1, +1, -1)$ at $\bold{X}$ and $(+1, -1, -1, +1)$ at $\bold{\Gamma}$ in order from the lowest band to the fourth band. Then, we get $[\bold{X}_{+1}^{(2)}]=\#\bold{X}_{+1}^{(2)}-\#\bold{\Gamma}_{+1}^{(2)}=0-1=-1$ for the lowest band only and $[\bold{X}_{+1}^{(2)}]=1-1=0$ for the first two bands together. Thus, according to Eq. (\ref{eq3a}), the lowest band carries a nonzero bulk polarization of $\bold{p}=\frac{1}{2}\left(\hat{\bold{x}}+\hat{\bold{y}}\right)$, but the first two bands together feature vanishing polarization $\bold{p}=\bold{0}$.

Figure \ref{fig.6.}(a) depicts the band structure of $h^{(4)}\left(|t|<|\lambda|;g=0\right)$ model along with $C^4$-rotation eigenvalues at $\bold{M}$ and at $\bold{\Gamma}$. At $\bold{M}$ and $\bold{\Gamma}$, the second and third bands are degenerate at zero energy. As these degenerate modes have different eigenvalues $\pm i$, there arises an ambiguity of whether we assign $+i$ or $-i$ to the $C^4$-rotation eigenvalue of the second band. This ambiguity, however, can be lifted up partially by the time-reversal symmetry, which enforces $[\bold{M}_{+i}^{(4)}]=[\bold{M}_{-i}^{(4)}]$, that we should choose the same values at $\bold{M}$ and at $\bold{\Gamma}$. Without loss of generality, $-i$ is assigned to the second band, see Fig. \ref{fig.6.}(a), which gives $[\bold{M}_{+i}^{(4)}]=1-1=0$ for the first two bands. Then, along with $[\bold{X}_{+1}^{(2)}]=0$ and $[\bold{M}_{+1}^{(4)}]=-1$, Equation \ref{eq3b} yields a half corner charge $Q_c = \frac{1}{2}$ for the first two bands.

Another way of interpreting this half charge is to consider this $C^4$-symmetric TCI model as a quadrupole insulator in a zero flux limit. A phase flux of $\theta$ can be achieved by complex tight-binding parameters $[t,\lambda]\rightarrow [t,\lambda]\times e^{+i\theta/4}$($\times e^{-i\theta/4}$) for hopping along(against) the direction of arrows illustrated in Fig. \ref{fig.6.}(b). Any finite phase flux upon a cyclic hopping opens a complete bandgap between the second and the third band, while maintaining the chiral symmetry \cite{HOTI_exp_5th}. In this setting, the $C^4$-rotation eigenvalue of the second band at $\bold{M}$, $+i$, is different from that at $\bold{\Gamma}$, $-i$, as shown in Fig. \ref{fig.6.}(c). While we cannot apply Eq. (\ref{eq3}) no longer as the time-reversal symmetry is broken due to the finite flux, the quadrupole moment $q_{xy}$ can be evaluated as
\begin{equation}
e^{i2\pi q_{xy}}=r_4^+(\bold{M})r_4^+(\bold{\Gamma})^*=r_4^-(\bold{M})r_4^-(\bold{\Gamma})^*, \label{eq4}
\end{equation}
where $r_4^\pm(\bold{k})$ is the $C^4$-rotation eigenvalue at $\bold{k}=\bold{M}/\bold{\Gamma}$ that satisfies $r_4^\pm(\bold{k})^2=\pm1$ \cite{Bena_PRB_2017, BoZhen}. From Fig. \ref{fig.5.}(c), we get $r_4^+(\bold{M})=-1$, $r_4^+(\bold{\Gamma})=+1$, $r_4^-(\bold{M})=+i$, $r_4^-(\bold{\Gamma})=-i$, and therefore $q_{xy}=Q_c=\frac{1}{2}$ \cite{Bena_PRB_2017}. 

Figure \ref{fig.6.}(c) shows the location of Wannier centers $\bf{p} = \frac{1}{2}\left(\hat{\bold{x}}+\hat{\bold{y}}\right)$ and the resulting charge distribution (in modulo unit charge) at quarter-filling (when the lowest energy band is filled) in the limit of $|t|\ll|\lambda|$. This quarter-filled configuration features edge charge density of $\frac{1}{2}$ per unit cell and FCA of $\phi=\frac{1}{4}$ \cite{FCA}. At half-filling (when the first two bands are filled), we have provided two different perspectives---(1) enforcing time-reversal symmetry or (2) treating the system as a time-reversal-broken quadrupole insulator with infinitesimal bandgap---that a corner charge index $Q_c=\frac{1}{2}$ can still be well-defined despite lack of a band gap at zero energy. The resulting charge distribution at half-filling (upto the second band) drawn in Fig. \ref{fig.6.}(d) shows a half corner charge and vanishing edge charge, as two overlapping Wannier centers from the first and second bands cancel the contribution to bulk polarization from each other.

\section{V. Role of chiral symmetry for ZCS in $C^4$ TCI with half corner charge}
Now that we have established a proper invariant $Q_c=\frac{1}{2}$ at half-filling, we investigate the crucial role of the chiral symmetry at half-filling for the existence of ZCSs. It is well studied in various systems \cite{HOTI_theory_2nd, HOTI_theory_3rd, HOTI_exp_2nd, HOTI_exp_3rd, HOTI_exp_4th, HOTI_exp_5th, Bena_PRB_2017, Bena_PRB_2019, SOTI_GSP} that the combination of a half fractional corner charge and the chiral symmetry guarantees a ZCS. If the bands below zero energy carry a half charge at a corner, the chiral symmetry ensures that the bands above zero energy also carry a half charge at the corner. Since the integration of local density of states over energy must be equal to the number of bands at each unit cell, the fractional corner charge in this case cannot be a charge surplus as it implies that the integration at the corner unit cell exceeds the number of bands. Thus, two half charge deficits, each from the lower and the upper bands, requires the existence of a corner state to compensate for total whole charge deficit, and this corner state should be pinned at zero energy due to the chiral symmetry.

The chiral symmetry in $C^4$-symmetric TCI is given as $S=diag[1,-1,1,-1]$ where the four sublattices are indexed in a clockwise order, and its presence $S h^{(4)} S^{-1} = -h^{(4)}$ gives rise to a band structure that is mirror-symmetric with respect to the zero energy as shown in Fig. \ref{fig.6.}(a)-(b). Since we have a half-corner charge and the chiral symmetry, a ZCS is expected to arise. Figure \ref{fig.6.}(e) shows that the expected ZCS is embedded in the bulk continuum due to absence of a bandgap at zero energy. In order to avoid numerical complication that the ZCS wavefunction gets generally mixed with other degenerate bulk states, we adopt the method used in Ref. \cite{C4_BIC}: as shown in Fig. \ref{fig.6.}(f), we introduced a uniform loss of $i\delta$ ($\delta=-0.05$) in the system except at small subsystems ($2\times2$ unit cell) at each corner. Then, the corner-localized states will be easily identified, as their imaginary part of eigenvalue becomes much smaller than other bulk modes, see Fig. \ref{fig.6.}(g). As expected, Figure \ref{fig.6.}(h) clearly shows the wavefunction of a truly corner-localized zero energy state at each corner.

\begin{figure}[t]
\centering
    \includegraphics[width=0.95\columnwidth]{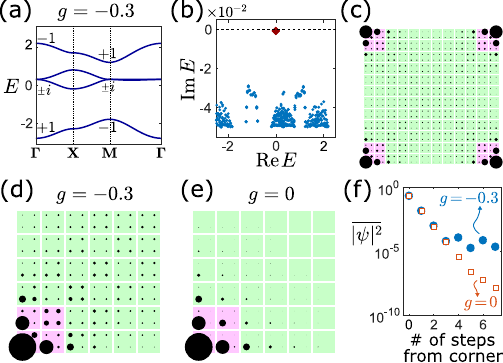}
    \caption{(a) Band structure of a chiral-symmetry-broken $h^{(4)}$ structure along with $C^4$-rotation eigenvalues at $\bold{\Gamma}$ and $\bold{M}$; $t=-0.25$, $\lambda=-1$, and $g=-0.3$. (b) Eigenvalues in a complex energy plane obtained by using the same setting described in Fig. \ref{fig.6.}(f). (c) Field profile of the modes marked as dark red dots in (b). (d) Zoom on one of the quadrants of (c). (e) Zoom on one of the quadrants of a ZCS wavefunction Fig. \ref{fig.6.}(h). (f) Average wavefunction amplitude $\overline{|\psi|^2}$ in a unit cell as a function of the number of grid steps ($x$+$y$) from the corner-most unit cell.}
    \label{fig.7.}
\end{figure}

The diagonal hopping across diagonally adjacent unit cells $g$, see Fig. \ref{fig.2.}(b), can be used to remove the chiral symmetry while preserving $C^4$ symmetry. Since $C^4$ symmtery is preserved, the perturbed structure with a finite $g$ still inherits the same $C^4$- and $C^2$-rotation eigenvalues for the modes at rotation-invariant momenta, given that $g$ is not too large to cause band inversion. In other words, a moderate strength of $g$ doesn't change bulk topological invariants $\bold{p}$ and $Q_c$ that are discussed in the previous section. Figure \ref{fig.7.}(a) shows that a finite $g$ breaks the chiral symmetry, as seen in the band structure that is not mirror-symmetric around the zero energy. Then, we observe that the modes that were ZCSs with $g=0$ now get hybridized with the bulk continuum due to broken chiral symmetry. A detailed analysis on how this hybridization occurs as a result of chiral symmetry breaking is provided in Ref. \cite{C4_BIC}. Figure \ref{fig.7.}(c)-(f) shows that the wavefunction amplitude of these hybridized modes remains finite in the bulk unlike the true ZCS wavefunction which decays exponentially from the corner. This observation verifies that the presence of the chiral symmetry with respect to zero energy plays a pivotal role in the existence of a ZCS, and therefore that the relevant topological indices for the ZCS should be investigated at half-filling instead at quarter-filling. For these reasons, we conclude that the bulk polarization of the first band does not play any role in the emergence of the ZCS in $C^4$-symmetric TCIs, contrary to the weak conjectures made in the previous works \cite{C4_BIC, C4_BIC2, FCA, C4_wrong_1st, C4_wrong_2nd}.

We note that the corner states observed in Ref. \cite{C4_wrong_1st,C4_wrong_2nd} are found in the bandgap between the first and the second band---e.g. around $E=-1$ in Fig. \ref{fig.6.}(a)---as they considered an embedded corner between the topological domain ($|t|<|\lambda|$) and the trivial domain ($|t|>|\lambda|$). However, these states lack any topological origin and don't share any common in their formation mechanism with the topological ZCS studied in this work and in Ref. \cite{C4_BIC, C4_BIC2}. In fact, these extra corner states are trivial defect states as a result of specific embedding condition. As these trivial corner states are not the main focus of our work, we leave the detailed discussion regarding these corner states in embedded structures to the Appendix.

\begin{figure}[t]
\centering
    \includegraphics[width=0.95\columnwidth]{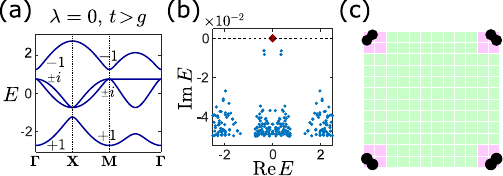}
    \caption{(a) Band structure along with $C^4$-rotation eigenvalues at $\bold{\Gamma}$ and $\bold{M}$ and (b) complex eigenvalues (from the same setting described in Fig. \ref{fig.6.}(f)) with $t=-1$, $g=-0.75$, and $\lambda=0$. (c) Field profile of the corner modes marked as dark red dots in (b)}
    \label{fig.8.}
\end{figure}

\section{VI. ZCS with no topological origin}
Lastly, we report a case where a ZCS arises at a corner of a system that is completely trivial both in bulk and in edge. In Fig. \ref{fig.8.}(a), we introduce another tight-binding parameter $s$ denoting for the diagonal hopping within a unit cell. Given $\lambda=0$ and $|t|>|g|$ in $h^{(4)}$, the resulting band structure in Fig. \ref{fig.8.}(a) features $[\bold{M}_{+1}^{(4)}]=[\bold{M}_{+i}^{(4)}]=[\bold{X}_{+1}^{(2)}]=0$; the $C^2$-rotation eigenvalues are $(+1, -1, -1, +1)$ both at $\bold{\Gamma}$ and at $\bold{X}$, and the $C^4$-rotation eigenvalues are $(+1, -i, +i, -1)$ both at $\bold{\Gamma}$ and at $\bold{M}$. Therefore, bulk polarization and corner charge vanish both at quarter-filling and at half-filling. These models also don't support any edge-localized modes in 1D edge dispersion of their nano-ribbon structures. Surprisingly, however, a ZCS is still present as shown in Fig. \ref{fig.8.}(b)-(c), clearly demonstrating that it is possible to obtain a corner defect state without any topological origin. In this sense, the present model $h^{(4)}(|t|>|g|; \lambda=0)$ serves as a pedagogical example that any corner states should not be assumed topological unless there is a physical causal relationship between the corner states and a certain topological invariant.

\section{VII. Conclusion}
In conclusion, we addressed that a topological correspondence between a corner state and a nontrivial bulk invariant should be claimed by a physical argument (e.g. a half charge with chiral symmetry), but not by coincidence of their existence conditions in terms of the Hamiltonian parameters. Our examples in $C^{3}$- and $C^4$-symmetric crystalline insulators clearly demonstrated that the bulk polarization and the corresponding fractional corner charge anomaly is not correlated with the emergence of the corner states at zero energy. In addition, we refined the bulk-corner correspondences for the corner states in these examples by identifying other topological invariants that are truly responsible for the corner state formation. We lastly showed that a corner state can appear even in a completely trivial insulator, which further strengthens our point that a corner-localized state may serve as an indicator of higher-order topology only when a solid bulk-corner correspondence precedes.

\section{acknowledgments}
This work was supported by the Office of Naval Research (ONR) under a Grant No. N00014-21-1-2056, and by the National Science Foundation (NSF) under the Grants No. DMR-1741788 and DMR-1719875. M.J. was also supported in part by the Kwanjeong Fellowship from Kwanjeong Educational Foundation.

\appendix
\section{Appendix}
\setcounter{equation}{0}
\setcounter{figure}{0}
\renewcommand{\theequation}{A\arabic{equation}}
\renewcommand{\thefigure}{A\arabic{figure}}
\subsection{1. Exact mapping of the edge-localized band in Figure \ref{fig.3.}(f) onto a chiral-symmetric SSH chain}
Here we show how the edge-localized band of a 1D-periodic nanoribbon structure out of a breathing Kagome lattice can be exactly mapped onto a chiral-symmetric 1D SSH model. We also explain why the chiral-partner band doesn't appear in the same band structure. 

The 1D SSH model is described by the following Hamiltonian:
\begin{equation}
\label{eqA1}
	\mathcal{H}_{\rm SSH} = \sum_{n\in \mathbb{Z}} \left(t\hat{c}_{A,n\bf{R}_{\rm 1}}^{\dagger}\hat{c}_{B,n\bf{R}_{\rm 1}}+\lambda\hat{c}_{A,n\bf{R}_{\rm 1}}^{\dagger}\hat{c}_{B,(n-1)\bf{R}_{\rm 1}}+c.c.\right), 
\end{equation}
where $\bf{R}_{\rm 1}$ is the primitive lattice vector, and $\hat{c}_{A/B,\bf{R}}$ and $\hat{c}_{A/B,\bf{R}}^{\dagger}$ is the annihilation and creation operators for the sublattice $A$/$B$ in the unit cell located at $\bf{R}$. By introducing the momentum space operators, $\hat{c}_{A/B,k_{1d}}=\frac{1}{\sqrt{L}}\sum_{n\in \mathbb{Z}} e^{-ink_{1d}}\hat{c}_{A/B,n\bf{R}_{\rm 1}}$ ($L$: the total length of SSH chain, $k_{1d}=\bf{k}\cdot\bf{R}_{\rm 1}$ where $\bf{k}$ is the Bloch momentum), we can obtain the momentum space Hamiltonian $H(k_{1d})$:
\begin{subequations}
\label{eqA2}
\begin{gather}
\mathcal{H}_{\rm SSH} = \sum_{k_{1d}\in [-\pi, \pi)} \begin{bmatrix} \hat{c}_{A,k_{1d}}^{\dagger} & \hat{c}_{B,k_{1d}}^{\dagger} \end{bmatrix}  H(k_{1d}) \it \begin{bmatrix} \hat{c}_{A,k_{1d}} \\ \hat{c}_{B,k_{1d}} \end{bmatrix}, \label{eqA2a}\\
H(k_{1d}) \it = \begin{bmatrix} 0 & t+\lambda e^{-i k_{1d}} \\ t+\lambda e^{i k_{1d}} & 0 \end{bmatrix}. \label{eqA2b}
\end{gather}
\end{subequations}
This Hamiltonian in Eq. (\ref{eqA2b}) is solved $H(k_{1d}) \vec{v}_{\pm} = E_{\pm} \vec{v}_{\pm}$ as below:
\begin{subequations}
\label{eqA3}
\begin{gather}
E_{\pm}(k_{1d})=\pm \rm{sign}\it (t) \sqrt{t^{\rm 2} + \lambda^{\rm 2} + \rm 2 \it t\lambda\cos(k_{\rm 1 \it d})}, \label{eqA3a}\\
\vec{v}_{\pm}(k_{1d})=\frac{1}{\sqrt{2}}\begin{bmatrix} \alpha_{\pm}(k_{1d}) & 1 \end{bmatrix}^{\dagger}, \label{eqA3b}\\
\alpha_{\pm}(k_{1d})=\pm \frac{\sqrt{|t|+|\lambda|e^{i k_{1d}}}}{\sqrt{|t|+|\lambda|e^{-i k_{1d}}}}. \label{eqA3c}
\end{gather}
\end{subequations}
Therefore, we get the following eigenbasis of $\mathcal{H}_{\rm SSH}$: $\mathcal{H}_{\rm SSH}\ket{k_{1d};\pm}=E_{\pm}(k_{1d})\ket{k_{1d};\pm}$, where
\begin{equation}
\label{eqA4}
\ket{k_{1d};\pm}=\frac{1}{\sqrt{2}}\left[\alpha_{\pm}(k_{1d})\hat{c}_{A,k_{1d}}^{\dagger}+\hat{c}_{B,k_{1d}}^{\dagger}\right]\ket{\rm vac}
\end{equation}

\begin{figure}[b]
\centering
    \includegraphics[width=0.95\columnwidth]{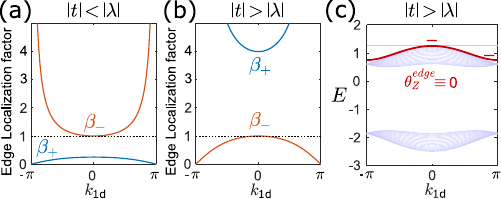}
    \caption{$\beta_{\pm}(k_{\text{1d}})$ in Eq. (\ref{eqA7}) calculated (a) with $t=-0.25$ and $\lambda=-1$ and (b) with $t=-1$ and $\lambda=-0.25$. (c) Edge dispersion with $t=-1$ and $\lambda=-0.25$ of nano-ribbon structure with an edge termination like one of the edges in Fig. \ref{fig.3.}(a); the edge localized band is colored red along with its inversion eigenvalues at high symmetry points.}
    \label{fig.A1.}
\end{figure}

Now, let us turn to the breathing Kagome lattice shown in Fig. \ref{fig.2.}(a). Consider edge-localized modes along an edge terminated by the side parallel to $\bf{R}_{\rm 1}$ drawn in Fig. \ref{fig.3.}(a). Let's label the two sublattices along the terminated edge as $A$ and $B$, and the other third sublattice as $C$. Then, the Hamiltonian for this edge-terminated Kagome lattice is given as:
\begin{equation}
\label{eqA5}
\begin{aligned}
	\mathcal{H}_{\rm Edge} = \sum_{n\in \mathbb{Z}, m\geq 0} ( & t\hat{c}_{A,n\bf{R}_{\rm 1}+\it m\bf{R}_{\rm 2}}^{\dagger}\hat{c}_{B,n\bf{R}_{\rm 1}+\it m\bf{R}_{\rm 2}}+\\
	& t\hat{c}_{B,n\bf{R}_{\rm 1}+\it m\bf{R}_{\rm 2}}^{\dagger}\hat{c}_{C,n\bf{R}_{\rm 1}+\it m\bf{R}_{\rm 2}}+ \\
	& t\hat{c}_{C,n\bf{R}_{\rm 1}+\it m\bf{R}_{\rm 2}}^{\dagger}\hat{c}_{A,n\bf{R}_{\rm 1}+\it m\bf{R}_{\rm 2}}+\\
	& \lambda\hat{c}_{A,n\bf{R}_{\rm 1}+\it m\bf{R}_{\rm 2}}^{\dagger}\hat{c}_{B,(n-1)\bf{R}_{\rm 1}+\it m\bf{R}_{\rm 2}}+ \\
	& \lambda\hat{c}_{B,n\bf{R}_{\rm 1}+\it (m+\rm 1)\bf{R}_{\rm 2}}^{\dagger}\hat{c}_{C,(n+1)\bf{R}_{\rm 1}+\it m\bf{R}_{\rm 2}}+\\
	& \lambda\hat{c}_{C,n\bf{R}_{\rm 1}+\it m\bf{R}_{\rm 2}}^{\dagger}\hat{c}_{A,n\bf{R}_{\rm 1}+\it (m+\rm 1)\bf{R}_{\rm 2}}+c.c.), 
\end{aligned}
\end{equation}
where $\bf{R}_{\rm 2}$ is the other primitive lattice vector that is not parallel to the terminated edge, see Fig. \ref{fig.3.}(a).
In order to map these edge-localized modes to 1D SSH eigenstates in Eq. (\ref{eqA4}), let's take the following ansatz:
\begin{subequations}
\label{eqA6}
\begin{gather}
\ket{k_{1d};\pm}=\left[\alpha_{\pm}(k_{1d})\hat{c}_{A,k_{1d};\pm}^{\dagger}+\hat{c}_{B,k_{1d};\pm}^{\dagger}\right]\ket{\rm vac}, \label{eqA6a}\\
\hat{c}_{A/B,k_{1d};\pm}=\sum_{n\in \mathbb{Z}, m\geq 0} \left[\beta_{\pm}(k_{1d})\right]^{m} e^{-ink_{1d}}\hat{c}_{A/B,n\bf{R}_{\rm 1}+\it m\bf{R}_{\rm 2}}. \label{eqA6b}
\end{gather}
\end{subequations}
Here, $\alpha_{\pm}(k_{1d})$ takes the same expression to Eq. (\ref{eqA3c}), and $\beta_{\pm}(k_{1d})$ signifies the edge localization. A proper normalization factor is not considered in Eqs. (\ref{eqA6}) for now, but this doesn't affect any of the following discussions. As we enforce $\mathcal{H}_{\rm Edge}\ket{k_{1d};\pm}=E_{\pm}(k_{1d})\ket{k_{1d};\pm}$, the wavefunction amplitudes on every sublattice $C$ are required to vanish and we obtain the following expression for the edge localization factor:
\begin{equation}
\label{eqA7}
\beta_{\pm}(k_{1d})=\frac{t}{\lambda} \frac{1+\alpha_{\pm}(k_{1d})}{1+\alpha_{\pm}(k_{1d})e^{-ik_{1d}}}.
\end{equation}
In order for the modes in Eq. (\ref{eqA6a}) to be truly edge-localized, the norm of $\beta$ should be less than 1. In fact, if we have $\lambda<t<0$ ($|t|<|\lambda|$) as in the main text, we get $|\beta_{+}(k_{1d})|<1\leq|\beta_{-}(k_{1d})|$, see Fig. \ref{fig.A1.}(a).

Therefore, the ansatz $\ket{k_{1d};+}$ is a valid eigenstate for $\mathcal{H}_{\rm Edge}$ with proper edge-localization, and this is the exact solution that describes the edge-localized band in Fig. \ref{fig.3.}(f) with dispersion relation of $E_{+}(k_{1d})$ from Eq. (\ref{eqA3a}). The chiral partner band $E_{-}(k_{1d})$ doesn't appear in the edge band dispersion, since $\ket{k_{1d};-}$ states violate the edge localization condition $|\beta_{-}(k_{1d})|\geq 1$. The same analysis can be repeated for more generic cases as depicted in Fig. \ref{fig.5.}(a), where all the hopping strengths $t/\lambda_{1,2,3}$ are different, and the edge localized band with a proper mapping onto 1D SSH chain can be found as long as $|t_i|<|\lambda_i|$ is met for each $i=1,2$. $t_3$ and $\lambda_3$ don't play any role in determining the existence of a ZCS at the corner made by edges along $\bf{R}_{\rm 1}$ and $\bf{R}_{\rm 2}$.

Lastly, we note that the breathing Kagome lattice with $|t|>|\lambda|$ (no bulk polarization) still supports an edge-localized band. Figure \ref{fig.A1.}(b) shows the edge localization factors for this trivial case $|t|>|\lambda|$; $|\beta_{-}(k_{1d})|\leq 1<|\beta_{+}(k_{1d})|$. Thus, in the same way, the ansatz $\ket{k_{1d};-}$ is a valid eigenstate for $\mathcal{H}_{\rm Edge}$ with proper edge-localization, and this is the exact solution that describes the edge-localized band in Fig. \ref{fig.A1.}(c) with dispersion relation of $E_{-}(k_{1d})$ from Eq. (\ref{eqA3a}). This edge-mapped SSH chain features a trivial (vanishing) Zak phase, as the inversion eigenvalues at $k_{\text{1d}}=0,\pi$ are equally $-1$.

\subsection{2. Trivial defect states in $C^4$ TCI at an embedded corner interfaced with a surrounding trivial domain}
As we briefly discussed at the end of the section V, there have been several works that studied the corner states in $C^4$ TCI at an embedded corner interfaced with a surrounding trivial domain \cite{C4_wrong_1st,C4_wrong_2nd}, where these corner states emerge in the bandgap between the first and the second bands instead at zero energy. These studies, without enough justification, conflated the origin of their corner state with that of the topological ZCS studied in this work. Here, however, we provide a detailed explanation on why the embedded corner states reported in Ref. \cite{C4_wrong_1st,C4_wrong_2nd} are trivial defect states, sharing no commonality in their formation mechanism with the ZCS studied in this work and in Ref. \citep{C4_BIC,C4_BIC2}.

\begin{figure*}
    \includegraphics[width=1.95\columnwidth]{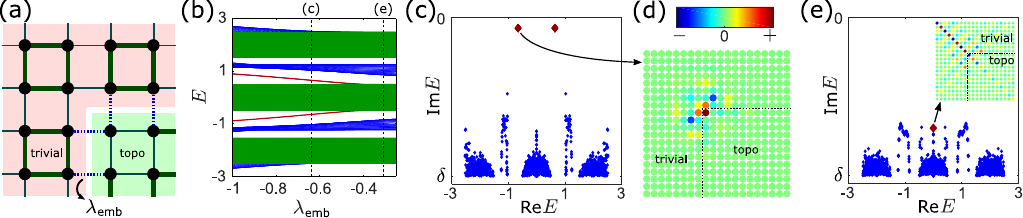}
    \caption{(a) An embedded corner between a topological domain and a trivial domain; $\lambda_{\text{emb}}$ refers to the coupling strength of the hopping across the domains. (b) Spectral flow, with varying $\lambda_{\text{emb}}$, of the embedded structure with a finite size (10 by 10 unit cells of the topological domain surrounded by the trivial domain of 5 unit-cell-long thickness); the green areas refer to bulk modes, the blue lines are edge(domain wall)-localized modes, and the dark red lines denote the corner localized states. (c) Complex eigenvalues (from the similar setting described in Fig. \ref{fig.6.}(f)) with $\lambda_{\text{emb}}=-0.64$. (d) Field profile of the embedded corner state. (e) Complex eigenvalues (from the similar setting described in Fig. \ref{fig.6.}(f)) with $\lambda_{\text{emb}}=-0.31$; the inset shows the field profile of the zero energy mode, which used to be the ZCS for the topological domain but lost the corner localization via hybridizing with the bulk modes in the trivial domain.}
    \label{fig.A2.}
\end{figure*}

Figure \ref{fig.A2.}(a) depicts the geometry of the topological domain ($|t|=0.25<|\lambda|=1$) interfacing with the trivial domain ($|t|=1>|\lambda|=0.25$) around an embedded corner. Naturally, the coupling strength of the hopping across the domains, $\lambda_{\text{emb}}$, would be given as a free parameter, which is determined by the microscopic details of the system and not by any topological effects. In the ring-resonator-based \cite{HOTI_exp_2nd} or circuit-based \cite{HOTI_exp_3rd} waveguide flatforms, the system can be designed for any arbitrary values of $\lambda_{\text{emb}}$. In the photonic crystal structures with subwavelength periodicities \cite{C4_wrong_1st,C4_wrong_2nd}, we can reasonably expect that the strength of $\lambda_{\text{emb}}$ would fall in the range between $\lambda$ in the trivial domain and $\lambda$ in the topological domain.

In Fig \ref{fig.A2.}(b), we computed the spectral flow of the embedded structure with a finite size, as $\lambda$ is varied between $-1$ and $-0.25$. The red curves show the embedded corner state predicted and observed in Ref. \cite{C4_wrong_1st,C4_wrong_2nd}. Figure \ref{fig.A2.}(c)-(d) depicts the corner localization and field profile of this corner state. It is clearly observed that, however, these states are not topologically protected in their spectral positions and get drifted as $\lambda_{\text{emb}}$ varies. Furthermore, in a certain range ($|\lambda_{\text{emb}}|<0.4$ in this example), these embedded corner states are lost, even though the crystalline symmetries of each bulk domain are not changing. In fact, as seen in the case $\lambda_{\text{emb}}=-0.31$ in Fig. \ref{fig.A2.}(e), there is no true corner state at all in the system (at least existing as a bound-in-continuum state like the ZCS in Fig. \ref{fig.6.}(g)-(h)).

Also, in the perspective of BBC, we have elaborated in the main text that there hasn't been reported any analytic or algebraic proof that the bulk polarization can be responsible for the existence of a corner state. Thus, even if the two domains exhibit different bulk polarization for the first band, we cannot conclude that these embedded corner states found in the band gap between the first and the second band are originating from the bulk polarization of the topological domain. Then, the immediate question is this: where did the original ZCS go? As the ZCS hybridizes via $\lambda_{\text{emb}}$ strongly with the bulk modes of the surrounding trivial domain, the corner localization is lost. The original ZCS will be restored again as we take $\lambda_{\text{emb}}\rightarrow 0$.

\end{document}